\documentclass{PoS}

\title{Study of {\boldmath $\Upsilon$}(5S) decays to B$^{\boldmath 0}$ and B$^{\boldmath +}$ mesons}

\ShortTitle{Study of $\Upsilon$(5S) decays to $B^0$ and $B^+$ mesons}

\author{\speaker{Alexey DRUTSKOY} \\
        (On behalf of the Belle collaboration) \\
        University of Cincinnati, Cincinnati, Ohio, US \\
        E-mail: \email{drutskoi@bpost.kek.jp}}

\abstract{
Decays of the $\Upsilon$(5S) resonance to channels with $B^+$ and $B^0$ mesons
are studied using a 23.6\,fb$^{-1}$ data sample
collected on the $\Upsilon$(5S) resonance with the Belle
detector at the KEKB asymmetric energy $e^+ e^-$ collider.
The fully reconstructed $B^+ \to J/\psi K^+$, $B^0 \to J/\psi K^{*0}$,
$B^+ \to \bar{D}^0 \pi^+$ and $B^0 \to D^- \pi^+$ decays are used
to obtain the $B^+$ and $B^0$ production rates per $b\bar{b}$ event,
$f(B^+) = (67.5 \pm 3.6 \pm 4.8)\%$ and
$f(B^0) = (70.4^{+5.2}_{-5.1} \pm 6.2)\%$.
Assuming equal rates to $B^+$ and $B^0$ mesons
in all channels produced at the $\Upsilon$(5S) energy,
we measure the fractions for $b\bar{b}$ event transitions to
the two-body and multi-body channels with $B^{+/0}$ meson pairs,
$f(B\bar{B}) = (5.1 \pm 0.9 \pm 0.4)\,\%$,
$f(B\bar{B}^*+B^*\bar{B}) = (12.6\,^{+1.2}_{-1.1} \pm 1.0)\,\%$,
$f(B^*\bar{B}^*) = (34.5\,^{+1.9}_{-1.8} \pm 2.7)\,\%$,
$f(B^{(*)}\bar{B}^{(*)}\pi(\pi)) = (16.4\,^{+1.6}_{-1.5} \pm 1.2)\,\%$,
$f(B\bar{B}\,\pi) = (0.0 \pm 1.1 \pm 0.2)\,\%$,
$f(B\bar{B}^\ast\pi+B^\ast\bar{B}\pi) = (6.8\,^{+2.1}_{-2.0} \pm 0.7)\,\%$,
and $f(B^\ast\bar{B}^\ast\pi) = (1.0\,^{+1.3}_{-1.2} \pm 0.3)\,\%$.}

\FullConference{The 2009 Europhysics Conference on High Energy Physics \\
		 16 - 22 July, 2009\\
		 Krakow, Poland}

\begin{document}

New aspects of beauty dynamics can be explored
using the large data sample recently collected by the Belle collaboration
at the energy of the $\Upsilon$(5S) resonance (also called
$\Upsilon$(10860)).
At this energy a $b\bar{b}$ quark pair can be produced and 
hadronized in various final states,
which can be classified as
two-body $B_s^0$ channels $B_s^0\bar{B}_s^0$, 
$B_s^0\bar{B}_s^\ast$, $B_s^\ast\bar{B}_s^0$, $B_s^\ast\bar{B}_s^\ast$,
two-body $B$ channels
$B\bar{B}$, $B\bar{B}^\ast$, $B^\ast\bar{B}$, $B^\ast\bar{B}^\ast$,
three-body channels $B\bar{B}\,\pi$, $B\bar{B}^\ast\,\pi$, 
$B^\ast\bar{B}\,\pi$, $B^\ast\bar{B}^\ast\,\pi$, and
four-body channel $B\bar{B}\,\pi \pi$.
Here $B$ denotes a $B^+$ or a $B^0$ meson and 
$\bar{B}$ denotes a $B^-$ or a $\bar{B}^0$ meson, and
the excited states decay to their ground states via
$B^\ast \to B\gamma$ and $B_s^\ast \to B_s^0\gamma$.
Fractions and decay parameters for all of these channels
provide important information about $b$-quark dynamics.

The first study of $B^+$ and $B^0$ mesons at the
$\Upsilon$(5S) was performed by the CLEO collaboration \cite{cleob}
using a $0.42$\,fb$^{-1}$ data sample,
where the fraction of events with $B^{+/0}$ meson pairs was found
to be $(58.9 \pm 10.0 \pm 9.2)\%$.
Only two-body $B$ meson channels were observed by CLEO.
Several theoretical papers are devoted to $\Upsilon$(5S) decays to final states
with the two-body $B_s^0$ and $B^{+/0}$ meson pairs \cite{teoa,teob,teoc,teod}
and with the three-body channels \cite{sim,lel}.
The three-body fractions were predicted to be about 
2-3 orders of magnitude smaller than two-body fractions.
Interesting information about $\Upsilon$(5S) behaviors can be obtained 
using multi-body decays \cite{est}.

In this analysis we use the data sample of $23.6\,\mathrm{fb}^{-1}$,
which was taken at the $\Upsilon$(5S) center-of-mass (CM) energy 
of $\sim$10867 MeV with the Belle detector \cite{belle} at KEKB \cite{kekb}.
The number of $b\bar{b}$ events in the data sample is
$N_{b\bar{b}}^{\Upsilon{\rm (5S)}} = (7.13 \pm 0.34) \times 10^6$.
We fully reconstruct the decay modes
$B^+ \to J/\psi K^+$, $B^0 \to J/\psi K^{*0}$,
$B^+ \to \bar{D}^0 \pi^+$ ($\bar{D}^0 \to K^+ \pi^-$ and $\bar{D}^0 \to K^+ \pi^+ \pi^- \pi^-$)
and $B^0 \to D^- \pi^+$ ($D^- \to K^+ \pi^- \pi^-$).
The $B$ decays are reconstructed and identified using two variables:
the energy difference $\Delta E\,=\,E^{CM}_{B}-E^{\rm CM}_{\rm beam}$
and the beam-energy-constrained mass
$M_{\rm bc} = \sqrt{(E^{\rm CM}_{\rm beam})^2\,-\,(p^{\rm CM}_{B})^2}$,
where $E^{\rm CM}_{B}$ and $p^{\rm CM}_{B}$ are the energy and momentum
of the $B$ candidate in the $e^+ e^-$ CM system,
and $E^{\rm CM}_{\rm beam}$ is the CM beam energy.
Two-body and multi-body channels with $B^{+/0}$ pairs are 
located in kinematically different regions of the $M_{\rm bc}$ and 
$\Delta E$ plane, however the central positions for all channels are
distributed in this plane along a straight line described approximately 
by the function \mbox{$\Delta E = m_B-M_{\rm bc}$}, where
$m_B$ is the nominal $B$ meson mass.

The two-dimensional $M_{\rm bc}$ and $\Delta E$ scatter plots
for the $B^+ \to J/\psi K^+$, $B^0 \to J/\psi K^{*0}$,
$B^+ \to \bar{D}^0 \pi^+$,
and $B^0 \to D^- \pi^+$ modes (with the following $J/\psi \to e^+ e^-$,
$J/\psi \to \mu^+ \mu^-$, $\bar{D}^0 \to K^+ \pi^-$,
$\bar{D}^0 \to K^+ \pi^+ \pi^- \pi^-$, and $D^- \to K^+ \pi^- \pi^-$ decays)
are shown in Fig.~1. Events are clearly
concentrated along the $B^{+/0}$ meson production line.

We used the inclined \mbox{$\Delta E+M_{\rm bc}-5.28$} projections of the
two-dimensional scatter plots
for all events within the range $5.268 < M_{\rm bc} < 5.43\,$GeV/$c^2$
to obtain integrated $B$ event yields.
To obtain the event yields we fit these distributions
with a function that includes two terms:
a Gaussian to describe the signal and a linear function
to describe background. 
Using the fit results, the average production rates
over studied $B$ modes, $f(B^+) = (67.5 \pm 3.6 \pm 4.8)\%$ and 
$f(B^0) = (70.4^{+5.2}_{-5.1} \pm 6.2)\%$, are obtained. 
The mean value over $B^+$ and $B^0$ modes is 
$f(B^{+/0}) = (68.5^{+3.0}_{-2.9} \pm 5.0)\%$.
Within errors this rate is in agreement
with the CLEO value of $f(B^{+/0}) = (58.9 \pm 10.0 \pm 9.2)\%$.

\begin{figure}
\vspace{-0.4cm}
\includegraphics[width=0.2\textwidth]{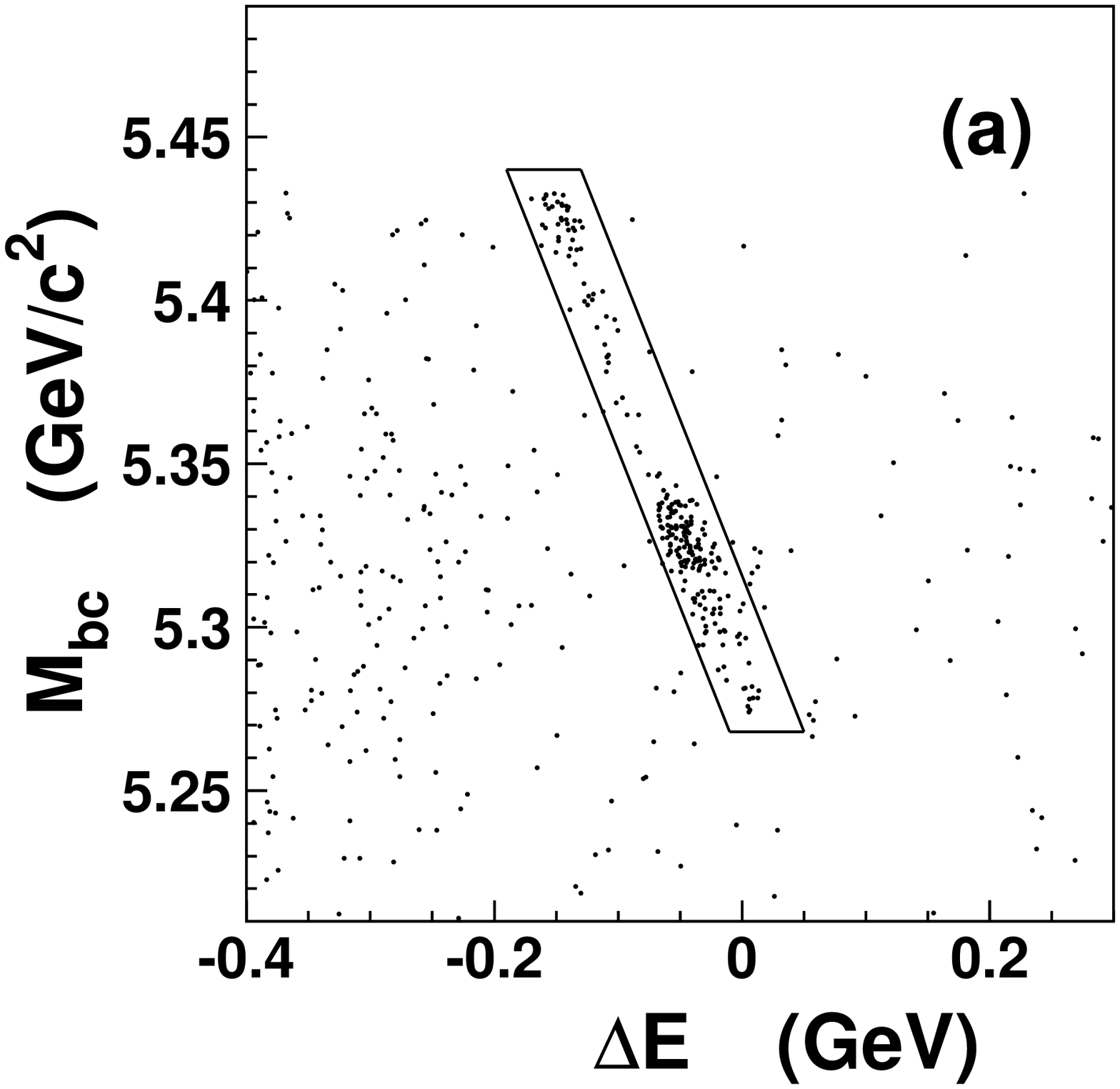}\includegraphics[width=0.2\textwidth]{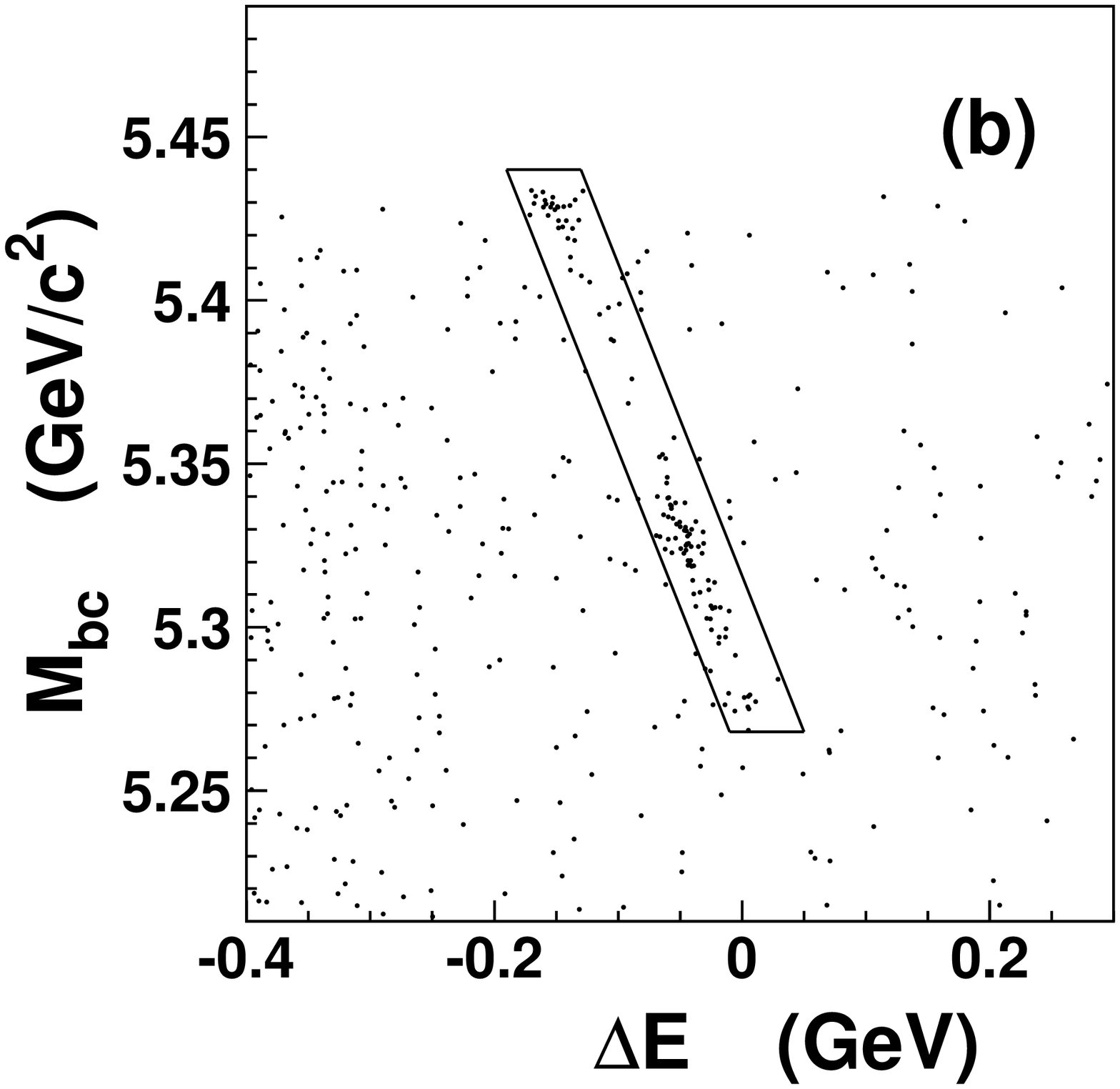}\includegraphics[width=0.2\textwidth]{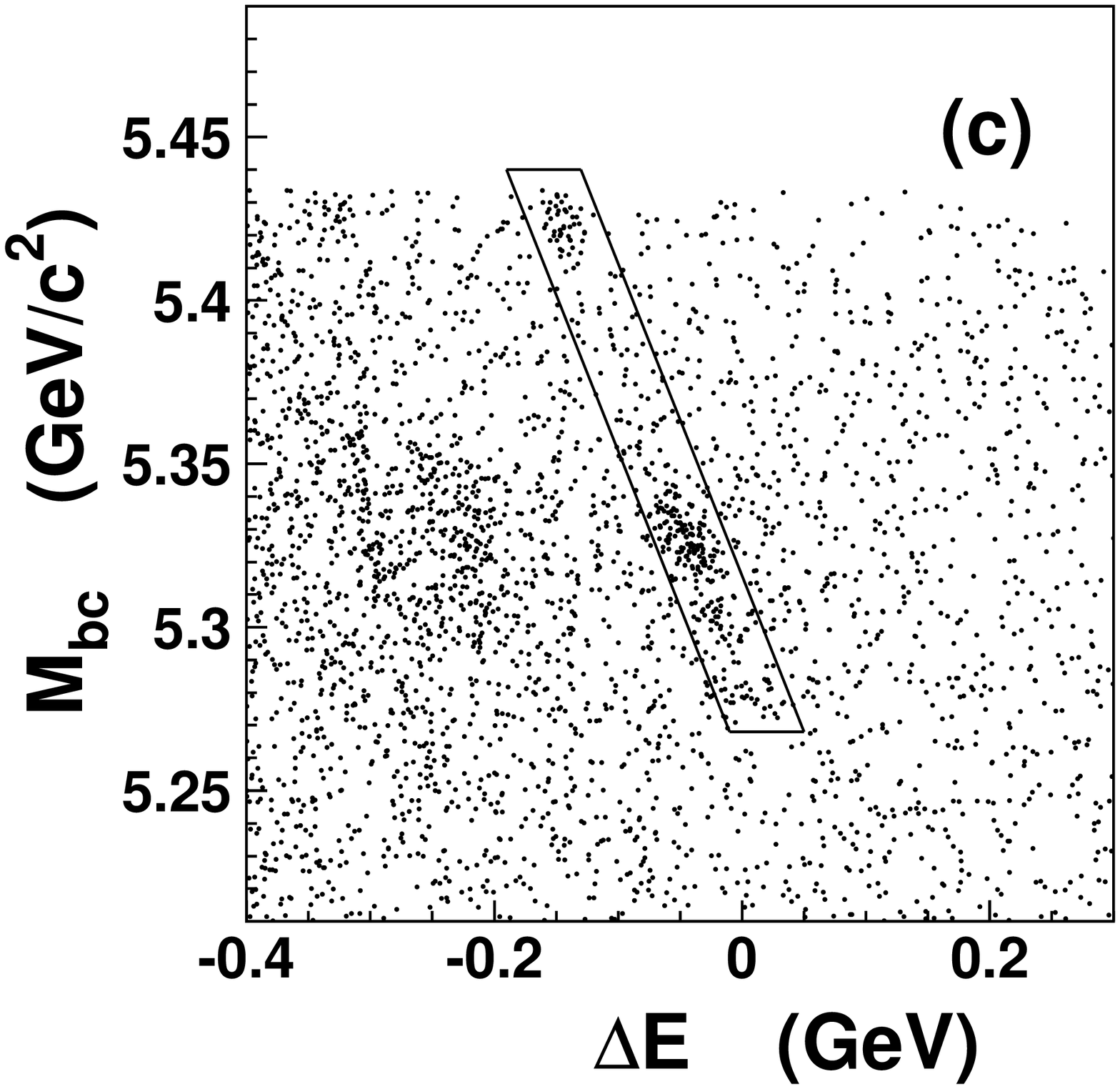}\includegraphics[width=0.2\textwidth]{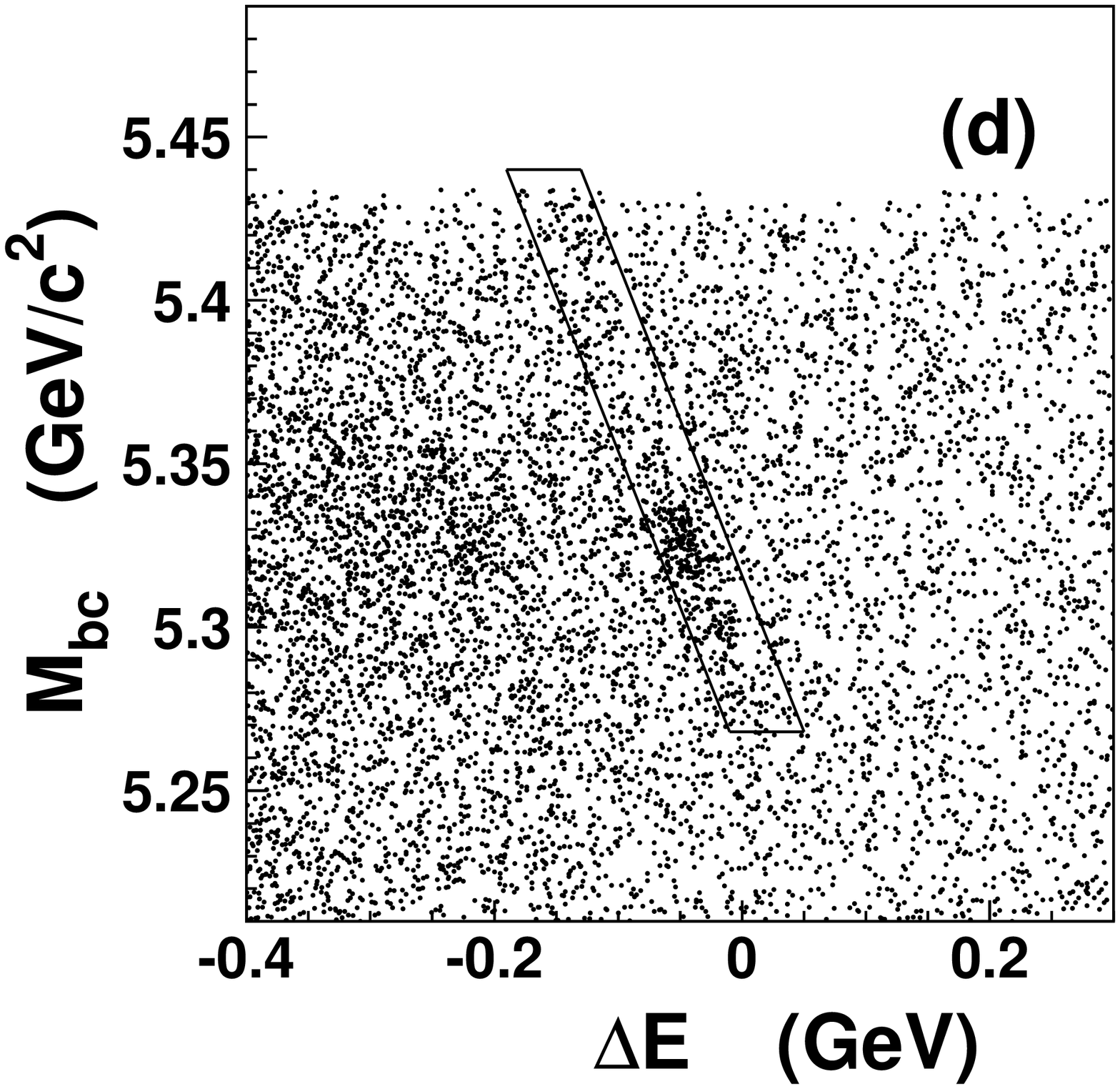}\includegraphics[width=0.2\textwidth]{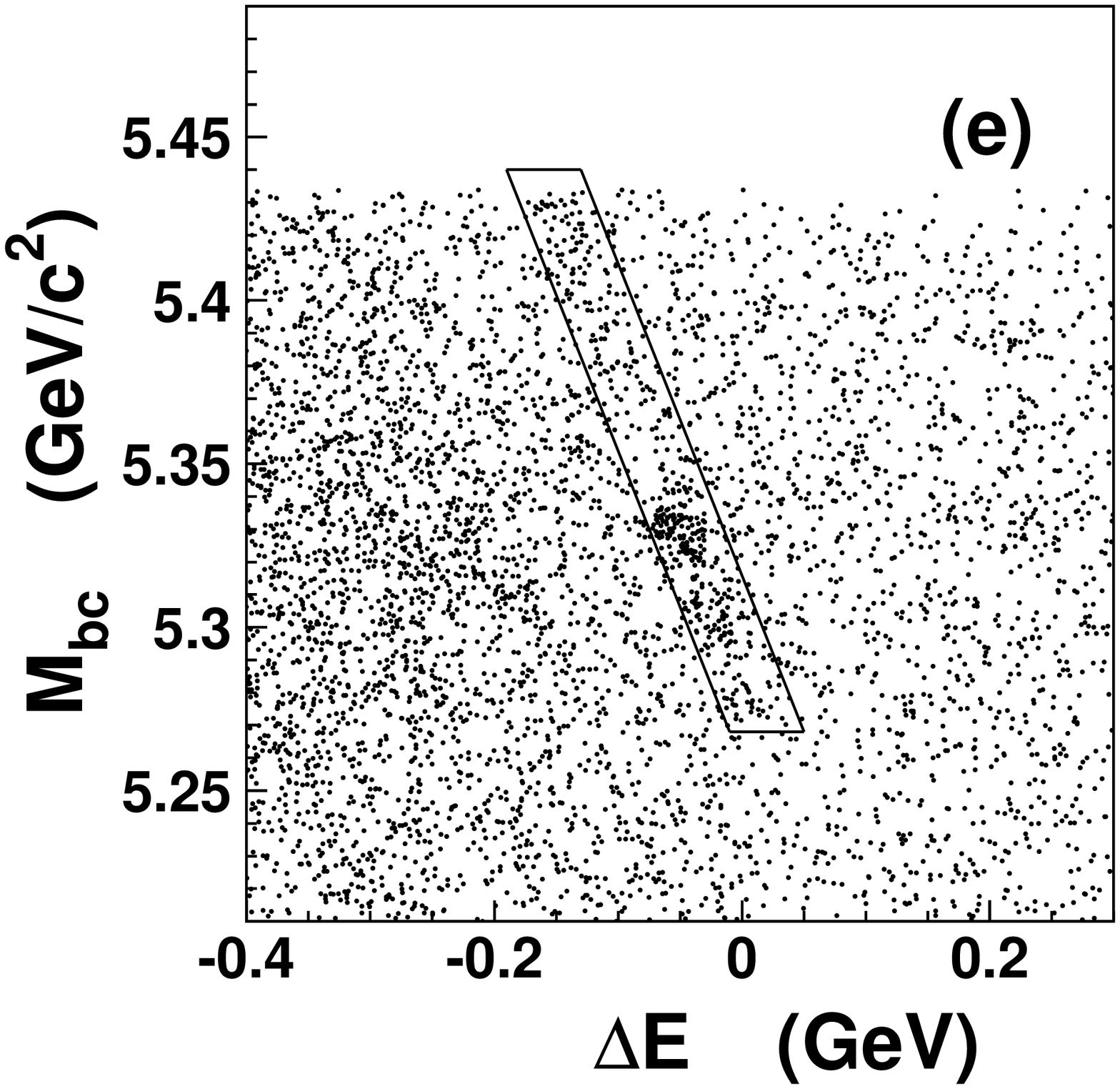}
\vspace{-0.3cm}
\caption{The $M_{\rm bc}$ and $\Delta E$ scatter plots
for the (a) $B^+ \to J/\psi K^+$, (b) $B^0 \to J/\psi K^{*0}$,
(c) $B^+ \to \bar{D}^0 (K^+ \pi^-) \pi^+$,
(d) $B^+ \to \bar{D}^0 (K^+ \pi^+ \pi^- \pi^-) \pi^+$,
and (e) $B^0 \to D^- \pi^+$ modes. The bands indicate the
signal regions corresponding
to the $5.268 < M_{\rm bc} < 5.44\,$GeV/$c^2$ and 
$|\Delta E+M_{\rm bc}-5.28| < 30\,$MeV intervals.}
\vspace{-0.4cm}
\label{fig1}
\end{figure}

Taking into account the obtained similarity of $f(B^+)$ and $f(B^0)$ rates,
it is natural to assume that the $B^+$ and $B^0$ mesons are symmetrically 
included in all possible channels with $B^{+/0}$ meson pairs.
Therefore the five studied $B$ decays are treated simultaneously
everywhere below.
First the $M_{\rm bc}$ projections of the two-dimensional
$M_{\rm bc}$ and $\Delta E$ distributions are studied.
Only events from the signal bands 
restricted to the $\pm 30\,$MeV interval in $\Delta E$ 
(which corresponds to a range of \mbox{(2.5-4.0)$\,\sigma$})
around the signal linear function (Fig.~1) are used to obtain the
signal $M_{\rm bc}$ distributions. To describe combinatorial backgrounds 
under the signals, we used left and right sideband regions corresponding
to a shift of 70~MeV in $\Delta E$ relative to the signal bands.

Fig.~2a shows the $M_{\rm bc}$ distributions for the specific two-, three- 
and four-body channels obtained from MC simulation for the 
$B^0 \to D^- \pi^+$ decay. 
Evidently the two-body channels are well separated 
from each other.
The decay matrix elements responsible
for the three- and four-body decays are not known and
the specific three- and four-body channel contributions
cannot be obtained in a model independent way from the fit of these 
$M_{\rm bc}$ distributions.
Therefore we restricted the fit procedure only 
to the region $5.268\,<\,M_{\rm bc}\,<$5.348~MeV/$c^2$
to extract the two-body channel fractions.
To obtain the sum of all three- and four-body channel fractions
we used the $5.348\,<\,M_{\rm bc}\,<$5.44~MeV/$c^2$ interval for
the fit procedure, similar to that used to obtain 
the full $f(B^{+/0})$ production rate for the entire interval
$5.268\,<\,M_{\rm bc}\,<$5.44~MeV/$c^2$, described above.
We measured the fractions of $b\bar{b}$ event transitions to
the two-body channels with $B^{+/0}$ meson pairs,
$f(B\bar{B}) = (5.1 \pm 0.9 \pm 0.4)\,\%$,
$f(B\bar{B}^*+B^*\bar{B}) = (12.6\,^{+1.2}_{-1.1} \pm 1.0)\,\%$,
$f(B^*\bar{B}^*) = (34.5\,^{+1.9}_{-1.8} \pm 2.7)\,\%$,
and to the sum of three-body and four-body channels with $B^{+/0}$ meson pairs,
$f(B^{(*)}\bar{B}^{(*)}\pi(\pi)) = (16.4\,^{+1.6}_{-1.5} \pm 1.2)\,\%$.
The sum of background subtracted $M_{\rm bc}$ distributions
for the five studied $B$ decays is shown in Fig.~2b, where the fit results
are also superimposed.

To obtain the three-body channel production rates we additionally 
used the charged pions directly produced in 
$B^{(*)}\bar{B}^{(*)}\,\pi^+$ channels. For all charged pions 
not contained in the reconstructed $B$ candidate, we
formed $B^{-/0}\,\pi^+$ combinations and calculated
the values $M_{\rm bc}^{\rm mis}$ and $\Delta E^{\rm mis}$ for the missing
$B$ meson.
To obtain these values we used the
energy and momentum of reconstructed $B\pi$ combination (all values 
in CM system): 
$E({B^{\rm mis}}) = 2 E_{\rm beam} - E({B\pi})$
and $p({B^{\rm mis}}) = p({B\pi})$.

\begin{figure}
\vspace{-0.4cm}
\begin{center}
\includegraphics[width=0.25\textwidth]{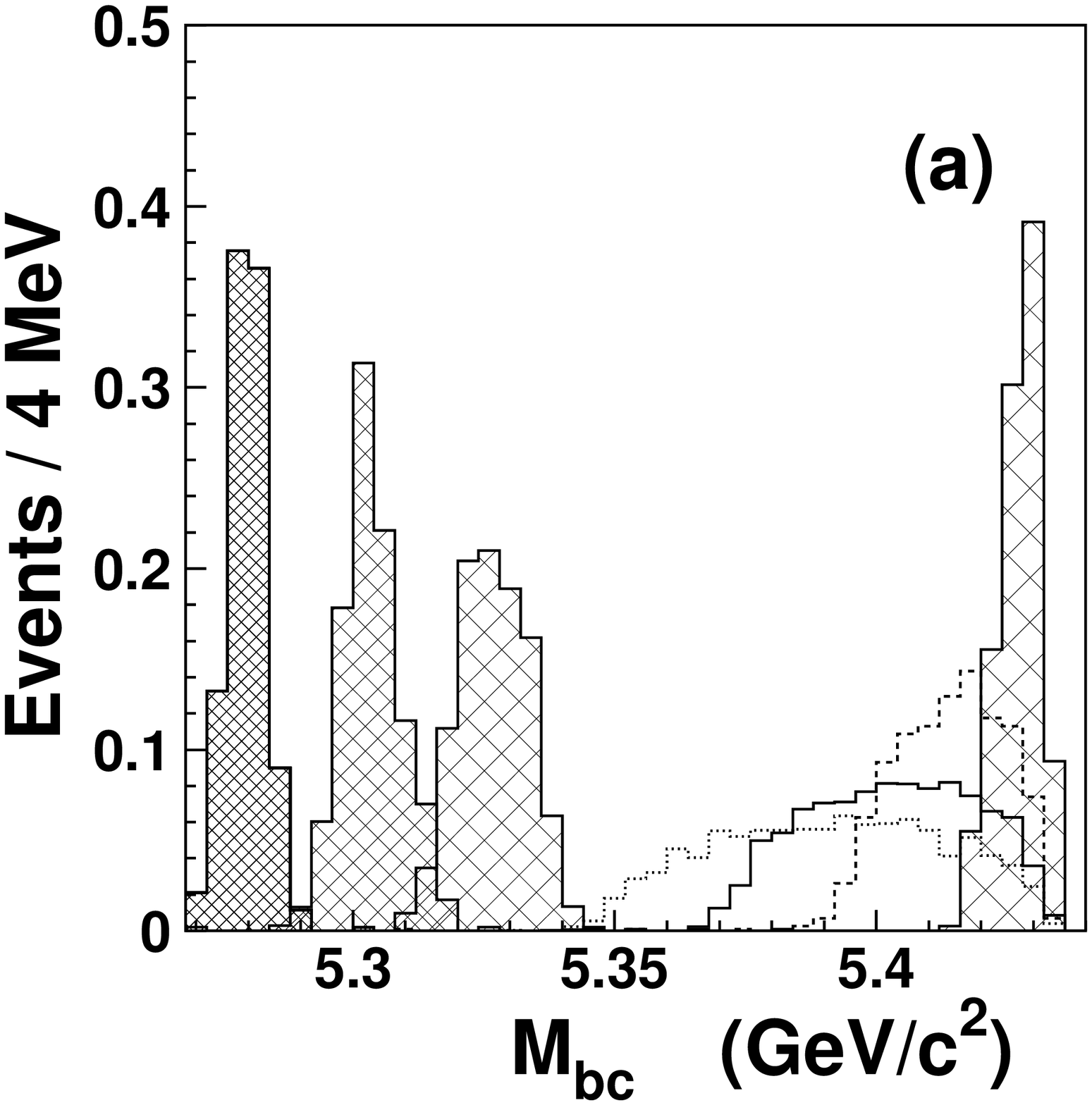}\includegraphics[width=0.25\textwidth]{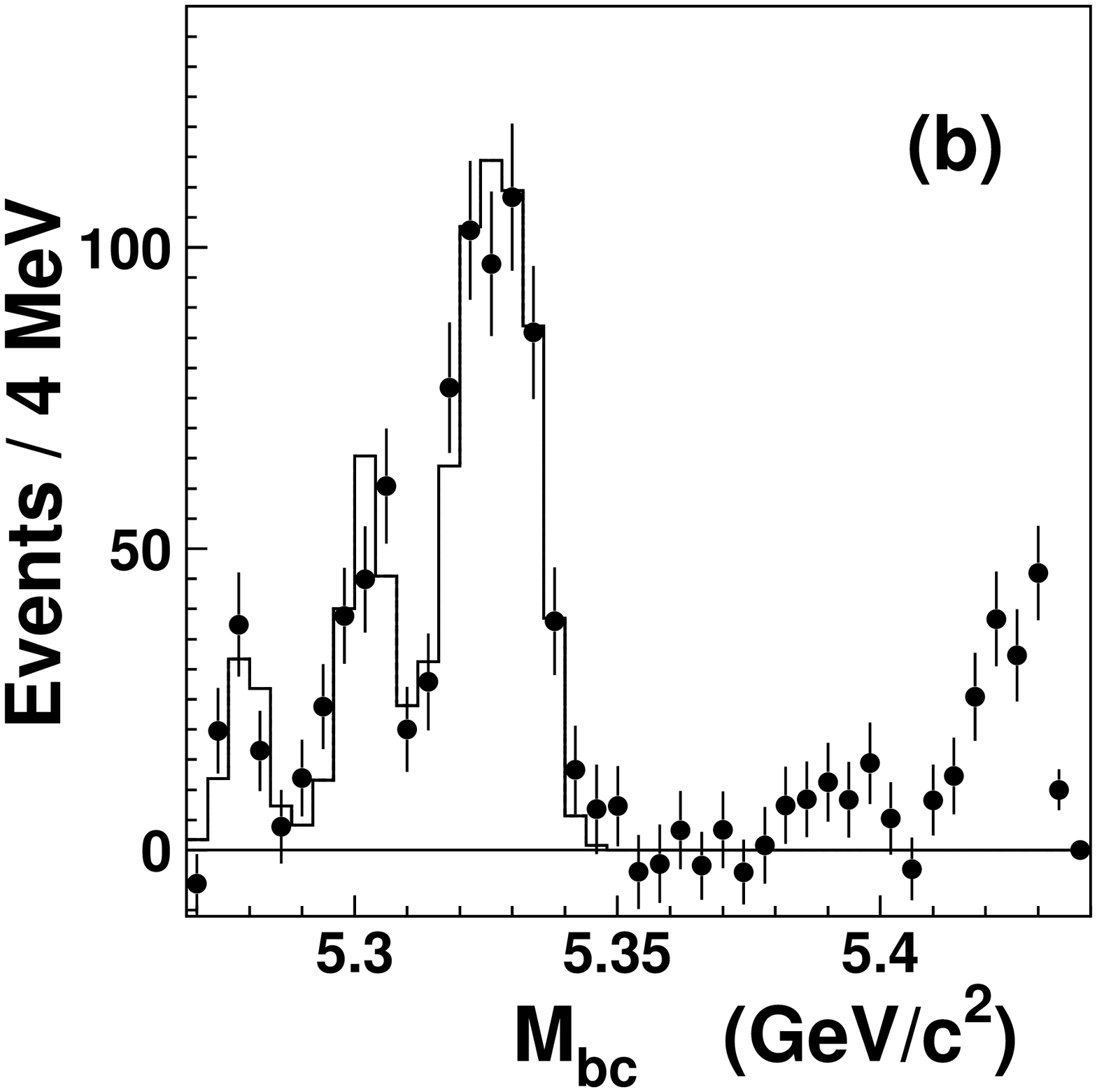}\includegraphics[width=0.25\textwidth]{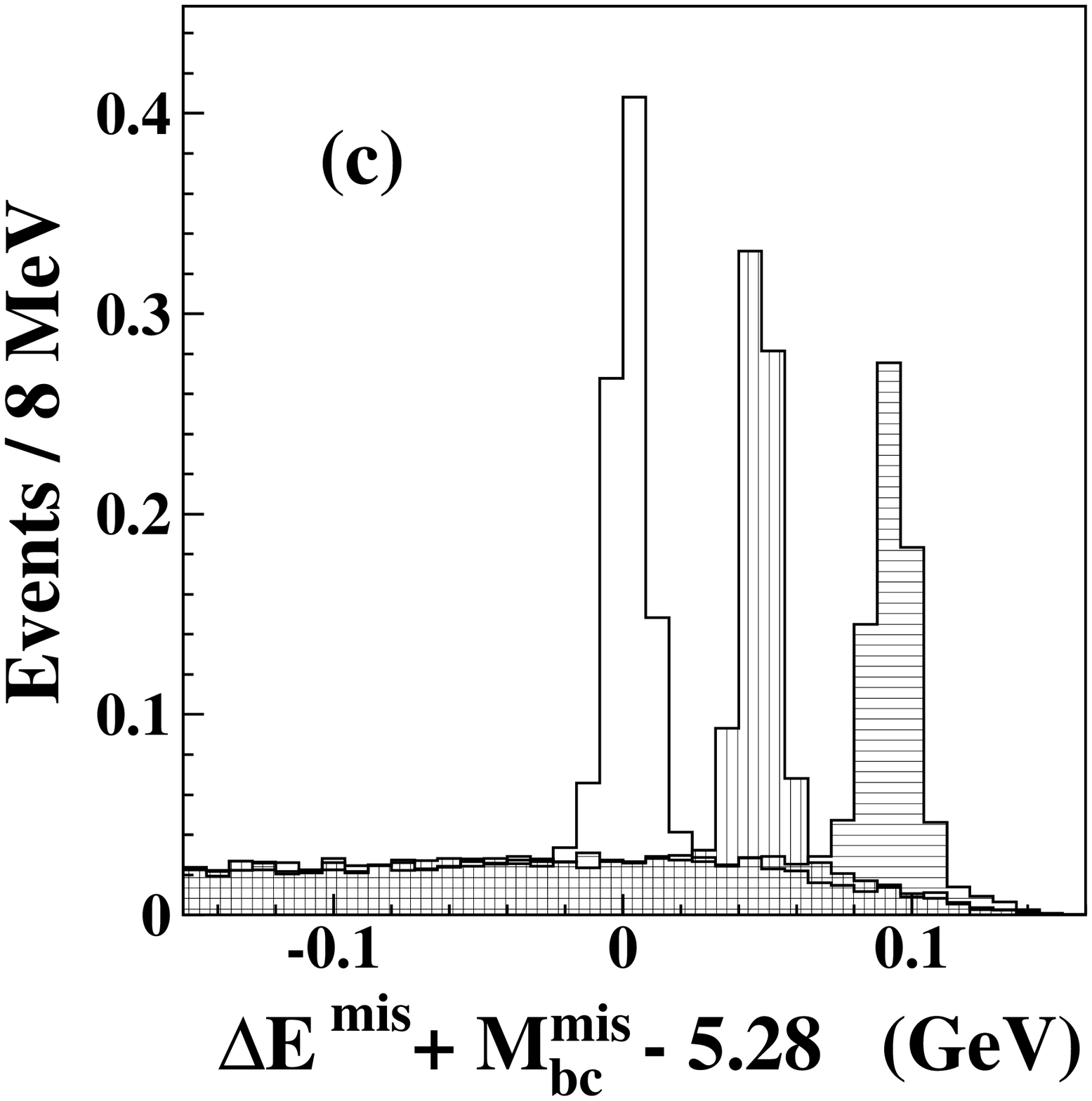}\includegraphics[width=0.25\textwidth]{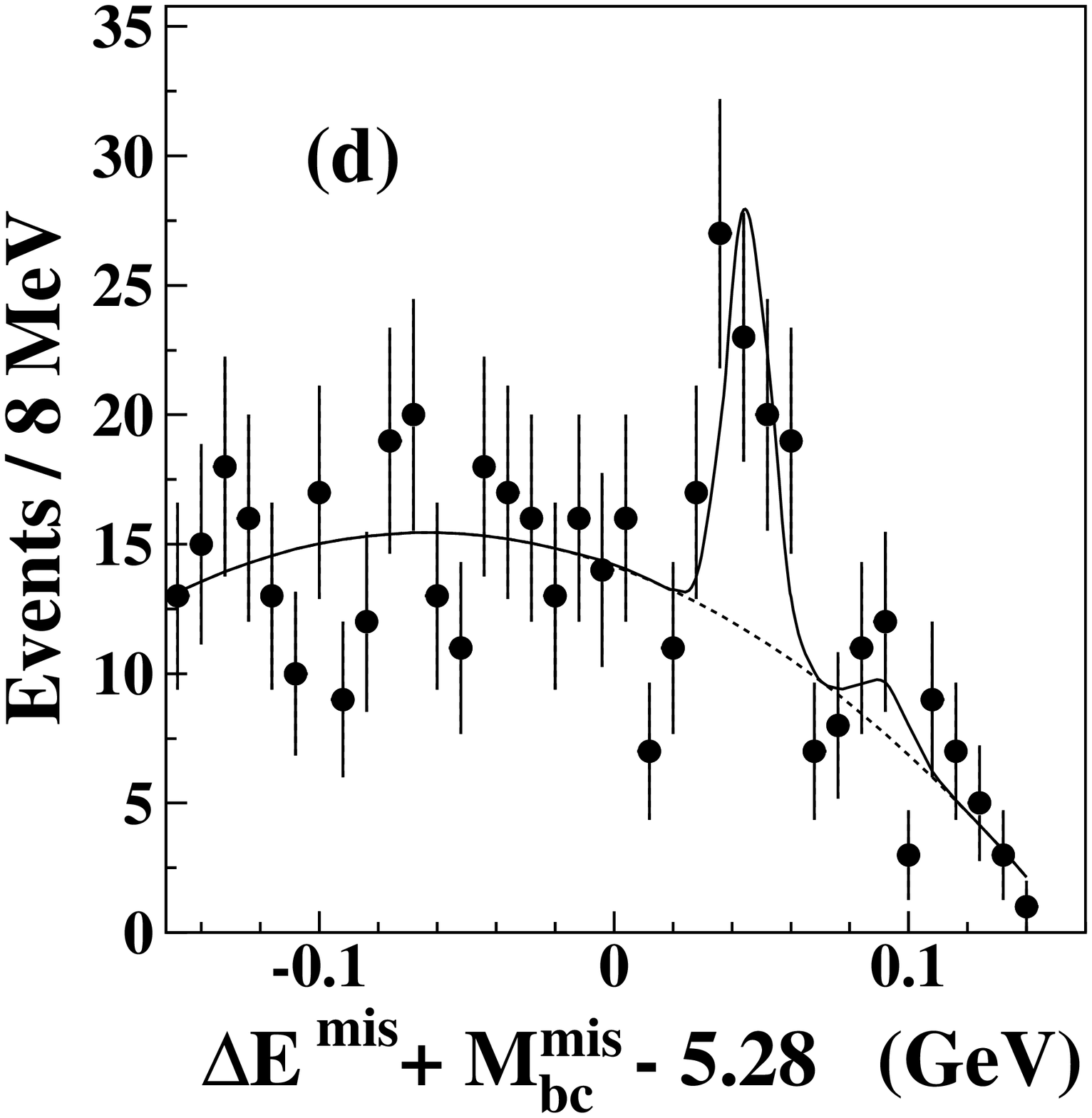}
\end{center}
\vspace{-0.6cm}
\caption{(a) MC simulated $M_{\rm bc}$ distributions for the 
$B^0 \to D^- \pi^+$ decay for 
(hatched histograms from the left to the right) (1) $B\bar{B}$, 
(2) $B\bar{B}^\ast+B^\ast\bar{B}$, (3) $B^\ast\bar{B}^\ast$ and
(4)  $B\bar{B}\,\pi \pi$ channels, and also for three-body
channels $B\bar{B}^\ast\,\pi+B^\ast\bar{B}\,\pi$ (histogram),
$B\bar{B}\,\pi$ (dotted histogram) and $B^\ast\bar{B}^\ast\,\pi$ (dashed
histogram). 
The distributions are normalized to unity. 
(b) $M_{\rm bc}$ data distribution after background subtraction
for the sum of five studied $B$ decays (points with error bars) and results of 
fit (histogram) used to extract two-body channel fractions.
\mbox{(c) The \mbox{$\Delta E^{\rm mis}+M_{\rm bc}^{\rm mis}-5.28$}}
distribution normalized per reconstructed $B$ meson
for the MC simulated $B^- \to J/\psi K^-$ decays
in the (peaks from the left to the right) $B\bar{B}\,\pi^+$,
$B\bar{B}^\ast\,\pi^+ +B^\ast\bar{B}\,\pi^+$, and 
$B^\ast\bar{B}^\ast\,\pi^+$
channels.
\mbox{(d) The \mbox{$\Delta E^{\rm mis}+M_{\rm bc}^{\rm mis}-5.28$} data}
distribution for $B^{-/0}\,\pi^+$ combinations 
for the sum of five studied $B$ modes. The curve shows the fit results described
in the text.}
\label{fig2}
\end{figure}


Figure 2c shows the corrected
\mbox{$\Delta E^{\rm mis}+M_{\rm bc}^{\rm mis}-5.28$}
projections ($m_B = 5.28\,$MeV/$c^2$) for MC simulated events where 
$B\bar{B}\,\pi^+$, $B\bar{B}^\ast\,\pi^+ +B^\ast\bar{B}\,\pi^+$, and
$B^\ast\bar{B}^\ast\,\pi^+$ channels 
are generated.
The reconstructed $B$ meson candidates are selected from the
signal region within the intervals $5.37 < M_{\rm bc} < 5.44\,$GeV/$c^2$ and
$|\Delta E + M_{\rm bc} - m_B| < 30\,$MeV.
To improve the resolution we applied additional correction requiring the exact
equality $\Delta E$ = $M_{\rm bc}-5.28$ for the reconstructed $B$ meson.
As we can see in Fig.~2c, the $B\bar{B}\,\pi^+$,
$B\bar{B}^\ast\,\pi^+ +B^\ast\bar{B}\,\pi^+$ and $B^\ast\bar{B}^\ast\,\pi^+$
channel contributions are well separated 
in \mbox{$\Delta E^{\rm mis}+M_{\rm bc}^{\rm mis}-5.28$}.

Finally the \mbox{$\Delta E^{\rm mis}+M_{\rm bc}^{\rm mis}-5.28$}
distribution is obtained in data for the sum of five reconstructed 
$B$ modes (Fig.~2d).
We fit the \mbox{$\Delta E^{\rm mis}+M_{\rm bc}^{\rm mis}-5.28$}
distribution with a function including four terms:
three Gaussians with fixed shapes and free normalizations used to describe 
the $B\bar{B}\,\pi^+$, $B\bar{B}^\ast\,\pi^+ +B^\ast\bar{B}\,\pi^+$,
and $B^\ast\bar{B}^\ast\,\pi^+$ channel contributions, and
a second order polynomial to describe background.
The central positions and widths of the Gaussians 
are obtained from fits of the MC simulated distributions
and are fixed in the fit to data.
Using the fit results we measured the fractions for $b\bar{b}$ 
event transitions to
the three-body channels with $B^{+/0}$ meson pairs,
$f(B\bar{B}\,\pi) = (0.0 \pm 1.1 \pm 0.2)\,\%$,
$f(B\bar{B}^\ast\pi+B^\ast\bar{B}\pi) = (6.8\,^{+2.1}_{-2.0} \pm 0.7)\,\%$,
and $f(B^\ast\bar{B}^\ast\pi) = (1.0\,^{+1.3}_{-1.2} \pm 0.3)\,\%$.
The three-body rates are calculated assuming the ratio of charged and neutral
directly produced pions to be 2:1 (Clebsch-Gordan coefficient).
It is interesting to note that using the method with the reconstructed direct
pions, we observe about one half of the rate obtained above for the
sum of the three-body and four-body channels.
This implies a sizable rate for the four-body $B\bar{B}\,\pi\pi$ channel,
however no definitive conclusion about this deficit can be reached
with the current statistics.

In conclusion, the production of $B^+$ and $B^0$ mesons is studied
at the $\Upsilon$(5S) energy.
\mbox{We measured} the average fractions per $b\bar{b}$ event
for integrated $B$ production, two-body, and multi-body channels.
The multi-body channels with unexpectedly large fractions
are observed for the first time.


\begin{thebibliography}{99}
\bibitem{cleob}
G. S. Huang {\it et al.} (CLEO Collaboration), 
Phys. Rev. D {\bf 75}, 012002 (2007).
%
%
%
\bibitem{teoa}
N. A. Tornqvist, Phys. Rev. Lett. {\bf 53}, 878 (1984).
%
\bibitem{teob}
S. Ono, A. I. Sanda, and N. A. Tornqvist, Phys. Rev. D\ {\bf 34}, 186 (1986).
%
\bibitem{teoc}
Yu. A. Simonov, and A. I. Veselov, Phys. Lett. B\ {\bf 671}, 55 (2009).
%
\bibitem{teod}
D. S. Hwang, and H. Son, arXiv:0812.4402 [hep-ph].
%
\bibitem{sim}
Yu. A. Simonov, and A. I. Veselov, JETP Lett. {\bf 88}, 79 (2008),
arXiv:0805.4518 [hep-ph].
%
\bibitem{lel}
L. Lellouch, L. Randall, and E. Sather, Nucl. Phys. B\ {\bf 405}, 55 (1993).
%
\bibitem{est}
I. J. General, S. R. Cotanch, and F. J. Llanes-Estrada,
Eur. Phys. J. {\bf C51}, 347 (2007).
%
\bibitem{belle}
Belle Collaboration, A. Abashian {\it et al.}, 
Nucl. Instr. and Meth. A\ {\bf 479}, 117 (2002).
%
\bibitem{kekb}
S. Kurokawa and E. Kikutani, Nucl. Instr. and Meth. A\ {\bf 499}, 1 (2003).
%
\end{thebibliography}
\end{document}